\documentclass{llncs}
\usepackage{epsf}
\usepackage{amsmath, graphicx}
\usepackage{multirow,multicol}
\usepackage{caption}
\usepackage{float}
\usepackage{latexsym}
\usepackage{amssymb,amsfonts}
\newcommand{\keywords}[1]{\par\addvspace\baselineskip
	\noindent\keywordname\enspace\ignorespaces#1}

\begin{document}

\title{Benchmarking Multimodal Sentiment Analysis}

\author{
Erik Cambria$^a$, Devamanyu Hazarika$^d$, Soujanya Poria$^c$, Amir Hussain$^b$, \\R.B.V. Subramaanyam$^d$}

\institute{
\footnotesize{$^a$ School of Computer Science and Engineering, NTU, Singapore\\
$^b$ School of Natural Sciences, University of Stirling, UK\\
$^c$ Temasek Laboratories, NTU, Singapore \\
$^d$ National Institute of Technology, Warangal, India}}

\maketitle

\begin{abstract}
We propose a framework for multimodal sentiment analysis and emotion recognition using convolutional neural network-based feature extraction from text and visual modalities. We obtain a performance improvement of 10\% over the state of the art by  combining visual, text and audio features. We also discuss some major issues frequently ignored in multimodal sentiment analysis research: the role of speaker-independent models, importance of the modalities and generalizability. The paper thus serve as a new benchmark for further research in multimodal sentiment analysis and also demonstrates the different facets of analysis to be considered while performing such tasks. 

\end{abstract}

\section{Introduction}

Emotion recognition and sentiment analysis have become a new trend in
social media, avidly helping users to understand opinions expressed
in social networks and user-generated content. With the advancement of communication technology, abundance of smartphones and the
rapid rise of social media, large amount of data is uploaded by the users as videos rather than text. For example, consumers tend to record their reviews and opinions on products using a web camera and
upload them on social media platforms such as YouTube or Facebook to inform subscribers of their views. These
videos often contain comparisons of products from competing brands, the pros and cons of product specifications, etc., which can aid prospective buyers in making an informed decision. 

The primary advantage of analyzing videos over textual
analysis for detecting emotions and sentiment from opinions is the surplus of behavioral cues. 
Video provides multimodal data in
terms of vocal and visual modalities. The vocal modulations and facial expressions in the visual data, along with textual
data, provide important cues to better identify true affective states of the opinion holder. Thus, a combination
of text and video data helps create a better emotion and sentiment analysis
model.

Recently, a number of approaches to multimodal sentiment analysis producing interesting results have been proposed~\cite{perez2013utterance,wollmer2013youtube,poria2015deep,zadeh2015micro}. However, there are major issues that remain unaddressed in this field, such as the role of speaker dependent and independent models, the impact of each modality across datasets, and generalization ability of a multimodal sentiment classifier. Not tackling these issues has presented difficulties in effective comparison of different multimodal sentiment analysis methods. 

In this paper, we address some of these issues and, in particular, propose a novel framework that outperforms the state of the art on benchmark datasets by more than 10\%. We use a deep convolutional neural network to extract features from visual and text modalities. 

The paper is organized as follows: In Section~\ref{sec:related} we give a brief literature review on multimodal sentiment analysis; in Section~\ref{sec:method} we present the method; experimental results and discussion are given in Section~\ref{exp}; finally, Section~\ref{sec:conclusion} concludes the paper.

\section{Related Work}

\label{sec:related}
Text-based sentiment analysis systems can be broadly categorized into knowledge-based and statistics-based systems \cite{camacsa}. While the use of knowledge bases was initially more popular for the identification of emotions and polarity in text, sentiment analysis researchers have recently been using statistics-based approaches, with a special focus on supervised statistical methods~\cite{pang2002thumbs,socher2013recursive}. 

In 1970, Ekman et al.~\cite{ekman1974universal} carried out extensive studies on facial expressions. Their research showed that universal facial expressions are able to provide sufficient clues to detect emotions. Recent studies on speech-based emotion analysis \cite{datcu2008semantic} have focused on identifying relevant acoustic features, such as fundamental frequency (pitch), intensity of utterance, bandwidth, and duration. 

As to fusing audio and visual modalities for emotion recognition, two of the early works were done by De Silva et al.~\cite{de1997facial} and Chen et al.~\cite{chen1998multimodal}. Both works showed that a bimodal system yielded a higher accuracy than any unimodal system. More recent research on audio-visual fusion for emotion recognition has been conducted at either feature level~\cite{kessous2010multimodal} or decision level~\cite{schuller2011recognizing}. 

While there are many research papers on audio-visual fusion for emotion recognition, only a few research works have been devoted to multimodal emotion or sentiment analysis using textual clues along with visual and audio modalities. Wollmer et al.~\cite{wollmer2013youtube} and Rozgic et al.~\cite{rozgic2012speech} fused information from audio, visual and textual modalities to extract emotion and sentiment. Metallinou et al.~\cite{metallinou2008audio} and Eyben et al.~\cite{eyben2010line} fused audio and textual modalities for emotion recognition. Both approaches relied on feature-level fusion. Wu et al.~\cite{wu2011emotion} fused audio and textual clues at decision level.

In this paper, we propose CNN-based framework for feature extraction from visual and text modality and a method for fusing them for multimodal sentiment analysis and emotion recognition. Our model outperforms the state of the art. In addition, we study the behavior of our method in the aspects rarely addressed by other authors, such as speaker independence, generalizability of the models and the performance of individual modalities.

\section{Method}
\label{sec:method}
\subsection{Textual Features}
\label{text}
\begin{figure}[h] 
	
	\centering    
	\includegraphics[height = 8 cm]{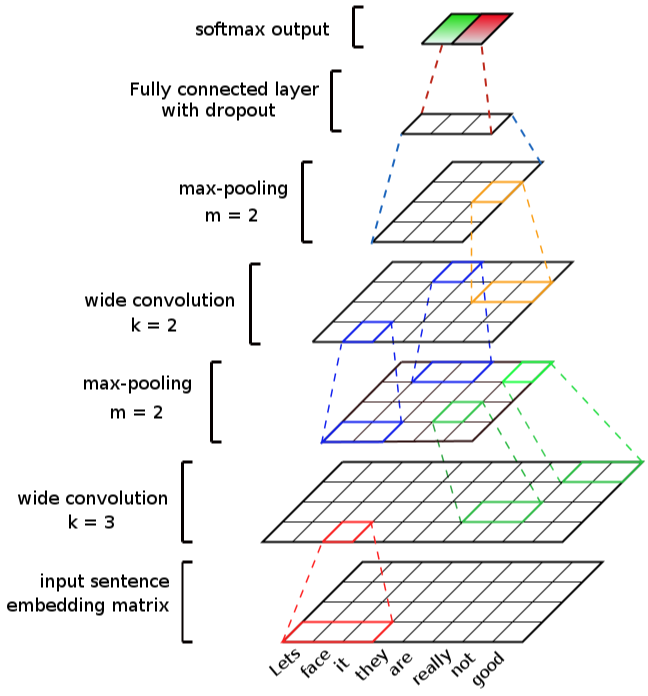}
	\caption[]{\small{CNN for feature extraction from text modality.}}
	\label{fig:textcnn}
	
\end{figure}

For feature extraction from textual data, we used a convolutional neural network (CNN). The trained CNN features were then fed into an SVM for classification, i.e., we used CNN as trainable feature extractor and SVM as a classifier. (see Figure~\ref{fig:textcnn}.)

The idea behind convolution is to take the dot product of a vector of $k$ weights $w_{k}$, known as kernel vector, with each $k$-gram in the sentence $s(t)$ to obtain another sequence of features $c(t)=(c_{1}(t),c_{2}(t),\ldots,c_{\mathrm{L}}(t))$: 

\begin{equation}
{c}_{j}={w}_{k}^{T}\cdot{\boldsymbol{\mathrm{x}}_{i:i+k-1}}.
\end{equation}

We then apply a max pooling operation over the feature map and take the maximum value $\hat{c}(t)=\mathrm{max}\{\boldsymbol{\mathrm{c}}(t)\}$ as the feature corresponding to this particular kernel vector. We used varying kernel vectors and window sizes to obtain multiple features. 

For each word $x_{i}(t)$ in the vocabulary, a $d$-dimensional vector representation, called word embedding, was given in a look-up table that had been learned from the data~\cite{mikolov2013efficient}. The vector representation of a sentence was a concatenation of the vectors for individual words. The convolution kernels are then applied to word vectors instead of individual words. Similarly, one can have look-up tables for features other than words if these features are deemed helpful. 

We used these features to train higher layers of the CNN to represent bigger groups of words in sentences. We denote the feature learned at a hidden neuron $h$ in layer $l$ as $F^{l}_{h}$. Multiple features are learned in parallel at the same CNN layer. The features learned at each layer are used to train the next layer:

\begin{equation}
F^{l}={\sum}_{h=1}^{n_{h}}w_{k}^{h}*F^{l-1},
\end{equation}

where * denotes convolution, $w_{k}$ is a weight kernel for hidden neuron $h$ and $n_{h}$ is the total number of hidden neurons.
The CNN sentence model preserves the order of words  by adopting convolution kernels of gradually increasing sizes, which span an increasing number of words and ultimately the entire sentence. 

Each word in a sentence was represented using word embeddings. We employed the publicly available word2vec vectors, which were trained on 100
billion words from Google News. The vectors were of dimensionality $d=300$, trained using the continuous
bag-of-words architecture ~\cite{mikolov2013efficient}. Words not present in the set of pre-trained words were
initialized randomly.

Each sentence was wrapped to a window of 50 words. Our CNN had two convolution layers. A kernel size of 3 and 4, each of them having 50 feature maps was used in the first convolution layer and a kernel size 2 and 100 feature maps in the second one. We used ReLU as the non-linear activation function of the network. The convolution layers were interleaved with pooling layers of dimension 2. 
We used the activation values of the 500-dimensional fully-connected layer of the network as our feature vector in the final fusion process. 

\subsection{Audio Features}
We automatically extracted audio features from each annotated segment of the videos. Audio features were also extracted in 30~Hz frame-rate; we used a sliding window of 100~ms. To compute the features, we used the open-source software openSMILE~\cite{eyben2010opensmile}. This toolkit automatically extracts pitch and voice intensity. Voice normalization was performed and voice intensity was thresholded to identify samples with and without voice. Z-standardization was used to perform voice normalization. 

The features extracted by openSMILE consist of several Low Level Descriptors (LLD) and their statistical functionals. Some of the functionals are amplitude mean, arithmetic mean, root quadratic mean, etc. Taking into account all functionals of each LLD, we obtained 6373 features.

\subsection{Visual Features}
Since, the video data is very large, we only considered every tenth frame in our training videos. The Constrained Local Model (CLM) was used to find the outline of the face in each frame \cite{baltruvsaitis20123d}. The cropped frame size was further reduced by scaling down to a lower resolution, thus creating our new frames for the video. In this way we could drastically reduce the amount of training video data. The frames were then passed through a CNN architecture similar to Figure~\ref{fig:textcnn}. 

\begin{figure}[h]
	\centering
	\includegraphics[scale=0.4]{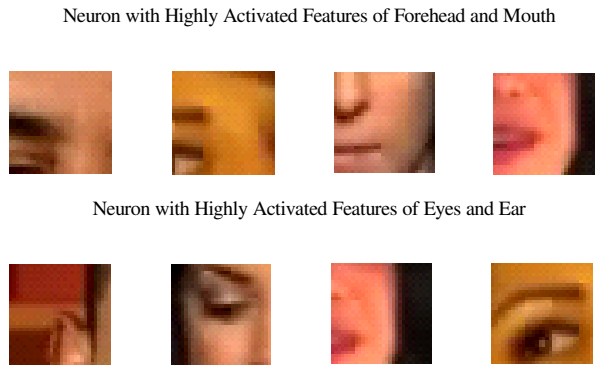}
	\caption{Top image segments activated at two feature detectors in the first layer of deep CNN}
	\label{fig:neu}
\end{figure}

A video comprised of a sequence of images. To capture the temporal dependence, we transformed each pair of consecutive images at $t$ and $t+1$ into a single image. And provided this transformed image as out input to the multilevel CNN. We used kernels of varying dimensions to learn Layer-1 2D features (shown in Figure \ref{fig:neu}) from the transformed input. Similarly, the second layer also used kernels of varying dimensions to learn 2D features. Down-sampling layer transformed features of different kernel sizes into uniform 2D features which was then followed by a logistic layer of neurons.

Thus, pre-processing involved scaling all video frames to half the resolution. Each pair of consecutive video frames were converted into a single frame to achieve temporal convolution features. All the frames were standardized to $250 \times 500$ pixels by padding with zeros. 

The first convolution layer contained 100 kernels of size 10$\times$20; the next convolution layer had 100 kernels of size $20\times30$; this layer was followed by a logistic layer of fully connected 300 neurons and a softmax layer. The convolution layers were interleaved with pooling layers of dimension \mbox{$2\times2$}. The activation of the neurons in the logistic layer were taken as the video features for the classification task.

\subsection{Fusion}
\begin{figure}[h]
	\centering
	\includegraphics[scale=0.5]{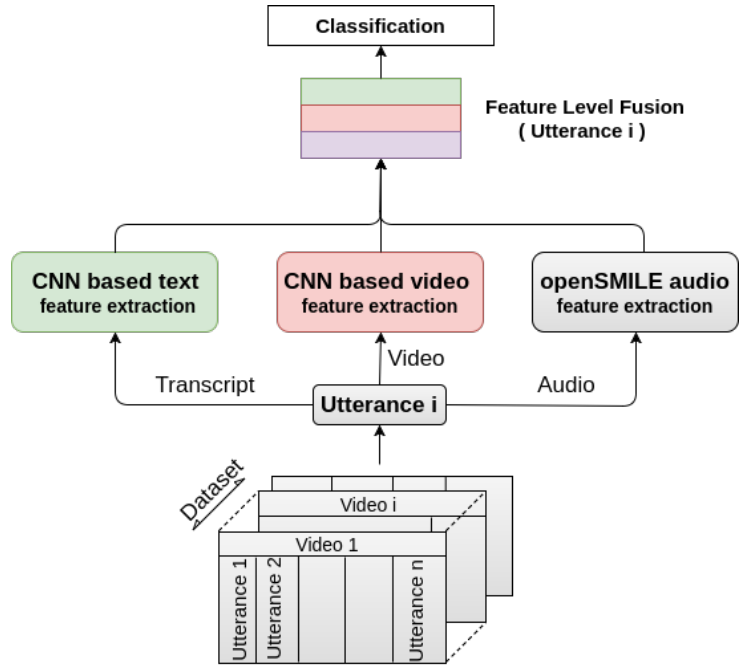}
	\caption{Overall architecture of the proposed method.}
	\label{fig:arch}
\end{figure}

In order to fuse the information extracted from each modality, we concatenated feature vectors extracted from each modality and sent the combined vector to a SVM for the final decision. This scheme of fusion is called feature-level fusion. Since the fusion involved concatenation and no overlapping merge or combination, scaling and normalization of the features were avoided. We discuss the results of this fusion in Section~\ref{exp}. The overall architecture of the proposed method can be seen in Figure~\ref{fig:arch}.

\section{Experiments and Observations}
\label{exp}
\subsection{Datasets}
\subsubsection{Multimodal Sentiment Analysis Datasets}
For our experiments, we use the MOUD dataset, developed by Perez-Rosas et al.~\cite{perez2013utterance}. They collected 80 product review and recommendation videos from YouTube. Each video was segmented into its utterances and each utterance was labeled by a sentiment (positive, negative and neutral). On average, each video has 6 utterances; each utterance is 5 seconds long. The dataset contains 498 utterances labeled either positive, negative or neutral. In our experiment we did not consider neutral labels, which led to the final dataset consisting of 448 utterances. We dropped the \emph{neutral} label to maintain consistency with previous work.

In a similar fashion, Zadeh et al.~\cite{zadeh2016multimodal} constructed a multimodal sentiment analysis dataset called Multimodal
Opinion-Level Sentiment Intensity (MOSI), which is bigger than MOUD, consisting of 2199 opinionated utterances, 93 videos by 89 speakers. The videos address a large array of
topics, such as movies, books, and products. In the experiment to address the generalizability issues, we trained a model on MOSI and tested on MOUD.

\subsubsection{Multimodal Emotion Recognition Dataset}
The USC IEMOCAP database \cite{busso2008iemocap} was collected for the purposes of studying multimodal expressive dyadic interactions. This dataset contains 12 hours of video data split into 5 minutes of dyadic interaction between professional male and female actors.  Each interaction session was split into spoken utterances. At least 3 annotators assigned to each utterance one emotion category: \emph{happy, sad, neutral, angry, surprised, excited, frustration, disgust, fear} and \emph{other}. In this work, we considered only the utterances with majority agreement (i.e., at least two out of three annotators labeled the same emotion) in the emotion classes of \emph{angry}, \emph{happy}, \emph{sad}, and \emph{neutral}. We take only these four classes so as to compare with the state-of-the-art~\cite{rozgic2012ensemble} and other authors.

All the mentioned datasets already contain manually created transcripts of the conversations or reviews. This might not be the case in a real-time video. However, with the availability of state of the art speech-to-text softwares, the task is efficient and trivial.




\subsection{Speaker-Independent Experiment}
Most of the research in multimodal sentiment analysis is performed on a datasets with speaker overlap in train and test splits. As we know, each individual is unique in his/her own way of expressing emotions and sentiments, finding generic, person independent features for sentimental analysis is very important. However, given this overlap, where the model has already seen the behaviour of a certain individual, the results do not scale to true generalization. In real world applications, the model should be robust to person variance. 

Thus, we performed person-independent experiments to emulate unseen conditions. This time, our train/test splits of the datasets were completely disjoint with respect to speakers. While testing, our models had to classify emotions and sentiments from utterances by speakers they have never seen before. Below, we enlist the procedure of this speaker-independent experiment:

\begin{itemize}
	\setlength\itemsep{1em}
	\item \textbf{IEMOCAP:} As this dataset contains 10 speakers, we performed a 10 fold speaker independent test, where in each round, one of the speaker was in the test set. The same SVM model was used as before and macro F\_score was used as a metric.
	
	\item \textbf{MOUD:} This dataset contains videos of about 80 people reviewing various products. Here, reviewers review products in Spanish. Each utterance in the video has been labeled to be either \emph{positive, negative or neutral}. In our experiments we consider only the \emph{positive and negative} sentiment labels. The speakers were divided into 5 groups and a 5-fold person independent experiment was run, where in every fold one out of the five group was in the test set. Finally we took average of the macro F\_score to summarize the results (see Table- \ref{tab:speakerindep} ).  
	
	\item \textbf{MOSI:} The MOSI dataset is a dataset rich in sentimental expressions where 93 people review topics in English. The videos are segmented with each segment's sentiment label scored between $+3$ to $-3$ by 5 annotators. We took the average of these labels as the sentiment polarity thus considering two classes \emph{positive and negative} as sentiment labels. Like MOUD, speakers were divided into 5 groups and a 5-fold person independent experiment was run. During each fold, around 75 people were in the train set and the remaining in the test set. The train set was further split randomly into 80\%--20\% and shuffled to generate train and validation splits for parameter tuning. 
\end{itemize}

\begin{table}[t]
	\centering
	\begin{tabular}{c|c|c|c|c}
		\hline
		\multirow{2}{*}{Modality }&\multirow{2}{*}{Source }
		& \multicolumn{3}{c}{}  \\
		& &IEMOCAP & MOUD & MOSI \\
		\hline
		\multirow{4}{*}{Unimodal}
		& A   & 51.52  & 53.70 & 57.14   \\
		& V   & 41.79 & 47.68 & 58.46   \\
		& T   & 65.13 & 48.40 & 75.16   \\
		\hline
		\multirow{3}{*}{Bimodal}
		& T + A   & 70.79 & 57.10 & 75.72   \\
		& T + V   & 68.55 & 49.22 & 75.06   \\
		& A + V   & 52.15 & 62.88 & 62.4   \\
		\hline
		\multirow{1}{*}{Multimodal}
		& T + A + V   & \textbf{71.59} & \textbf{67.90} & \textbf{76.66}   \\
		\hline
	\end{tabular}
	\caption{\textbf{Speaker Independent: }Macro F\_score reported for speaker independent classification. \emph{IEMOCAP:} 10-fold speaker independent average. \emph{MOUD:} 5-fold speaker independent average. \emph{MOSI:} 5-fold speaker independent average. \emph{Notes:} A stands for Audio, V for Video, T for Text.}
	\label{tab:speakerindep}
\end{table}

\begin{table}
	\newcommand\Y{\hphantom{$^1$}}
	\newcommand\X{\%\Y}
	\centering
	\begin{tabular}{c|c|c|c}
		\hline
		Modality & Source & IEMOCAP & MOSI \\
		\hline
		\multirow{3}{*}{Unimodal}
		& Audio   & 66.20 & 64.00  \\
		& Video   & 60.30 & 62.11 \\
		& Text   & 67.90 & 78.00 \\
		\hline
		\multirow{3}{*}{Bimodal}
		& Text + Audio   & 78.20 & 76.60\\
		& Text + Video    & 76.30 & 78.80\\
		& Audio + Video   & 73.90 & 66.65\\
		\hline
		\multirow{1}{*}{Multimodal}
		& Text + Audio + Video  & \textbf{81.70} & \textbf{78.80}\\
		& Text + Audio + Video  & \textbf{69.35}$^1$ & \textbf{73.55}$^2$\\
		\hline
	\end{tabular}
	\caption{\textbf{Speaker Dependent: }Ten-fold cross-validation results on IEMOCAP dataset and 5-fold CV results (macro F\_Score) on MOSI dataset. $^1$By \cite{rozgic2012ensemble}; $^2$by \cite{poria2015deep}.}
	\label{tab:speakerdep}
\end{table}

\subsubsection{Comparison with the Speaker Dependent Experiment}

\begin{figure}[H]
	\centering
	\includegraphics[width=0.45\linewidth]{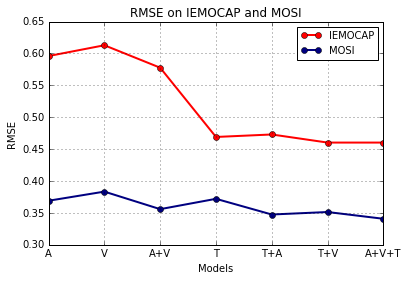}%
	\includegraphics[width=0.45\linewidth]{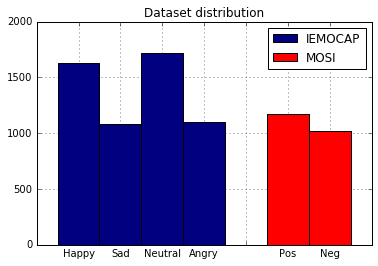}
	\includegraphics[width=0.45\linewidth]{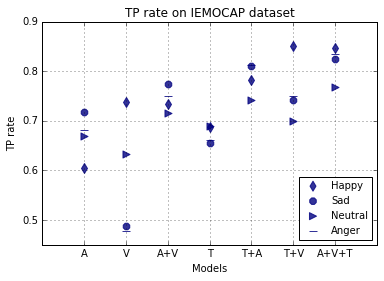}
	\includegraphics[width=0.45\linewidth]{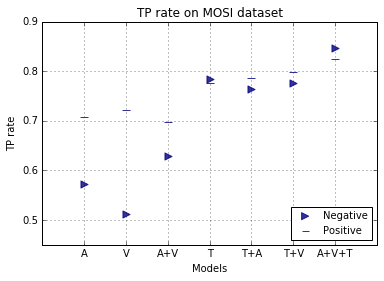}
	
	\caption{Experiments on IEMOCAP and MOSI datasets. Top left figure shows the Root Mean Square Error (RMSE) of the models on IEMOCAP and MOSI. Top right figure shows the dataset distribution. Bottom left and bottom right figure present TP-rate on of the models on IEMOCAP and MOSI dataset respectively. }
	\label{fig:fig1}
\end{figure}

In comparison to speaker dependent experiment, speaker independent experiment performs poor. This is due to the lack of knowledge about speakers in the dataset. Table \ref{tab:speakerdep} shows the performance obtained in the speaker dependent experiment. It can be seen that audio modality consistently performs better than visual modality in both MOSI and IEMOCAP datasets. The text modality plays the most important role in both emotion recognition and sentiment analysis. The fusion of the modalities show more impact for emotion recognition than on sentiment analysis. RMSE and TP-rate of the experiments using different modalities on IEMOCAP and MOSI datasets are shown in Figure \ref{fig:fig1}.

\subsection{Contributions of the Modalities}
\begin{figure}[t]
	\centering
	\includegraphics[scale=0.5]{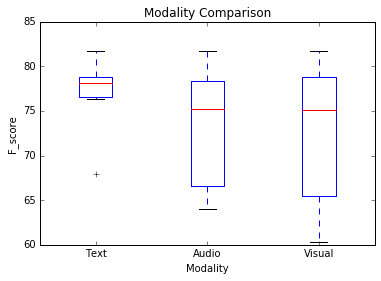}%
	\caption{Performance of the modalities on the datasets. Red line indicates the median of the F\_score.}
	\label{fig:fig2}
\end{figure}

As expected in all kinds of experiments, bimodal and trimodal models have performed better than unimodal models. Overall, audio modality has performed better than visual on all the datasets. Except the MOUD dataset, the unimodal performance of text modality is notably better than other two modalities; see Figure \ref{fig:fig2}. Table \ref{tab:speakerdep} also presents the comparison with state of the art. The present method outperformed state of the art by 12\% and 5\% respectively on the IEMOCAP and MOSI datasets.\footnote{We have reimplemented the method by Poria et al.~\cite{poria2015deep}} The method proposed by Poria et al.\ is similar to us except they used a standard CLM based facial feature extraction method. So, our proposed CNN based visual feature extraction algorithm has helped to outperform the method by Poria et al.

\begin{table}[H]
	\centering
	\begin{tabular}{c|c|c}
		\hline
		Modality & Source  & Macro F\_Score \\
		\hline
		\multirow{4}{*}{Unimodal}
		& Audio & 41.60 $\%$ \\
		& Video & 45.50 $\%$ \\
		& Text & 50.89 $\%$ \\
		\hline
		\multirow{3}{*}{Bimodal}
		& Text + Audio & 51.70 $\%$ \\
		& Text + Video & 52.12 $\%$ \\
		& Audio + Video & 46.35 $\%$ \\
		\hline
		\multirow{1}{*}{Multimodal}
		& Text + Audio + Video & 52.44 $\%$ \\
		\hline
	\end{tabular}
	\caption{\textbf{Cross dataset results: }Model (with previous configurations) trained on MOSI dataset and tested on MOUD dataset.}
	\label{tab:gen}
\end{table}

\subsection{Generalizability of the Models}
To test the generalization ability of the models we have trained framework on MOSI dataset in speaker independent fashion and tested on MOUD dataset. From Table \ref{tab:gen} we can see that the trained model on MOSI dataset performed poorly on MOUD dataset. While harvesting the reason for it, we have found mainly two major issues. First, reviews in MOUD dataset had been recorded in Spanish so audio modality miserably fail in recognition as MOSI dataset contains reviews in English. Second, text modality has performed very poor, too, for the same reason. A more comprehensive study would be to perform generalizability tests on datasets in the same language. However we were unable to do this owing to the lack of benchmark datasets.

Also, similar experiments of cross dataset generalization was not performed on emotion detection given the availability of only a single dataset - IEMOCAP.

\subsection{Visualization of the Datasets}
\begin{figure}[H]
	\centering
	\includegraphics[width=0.5\linewidth]{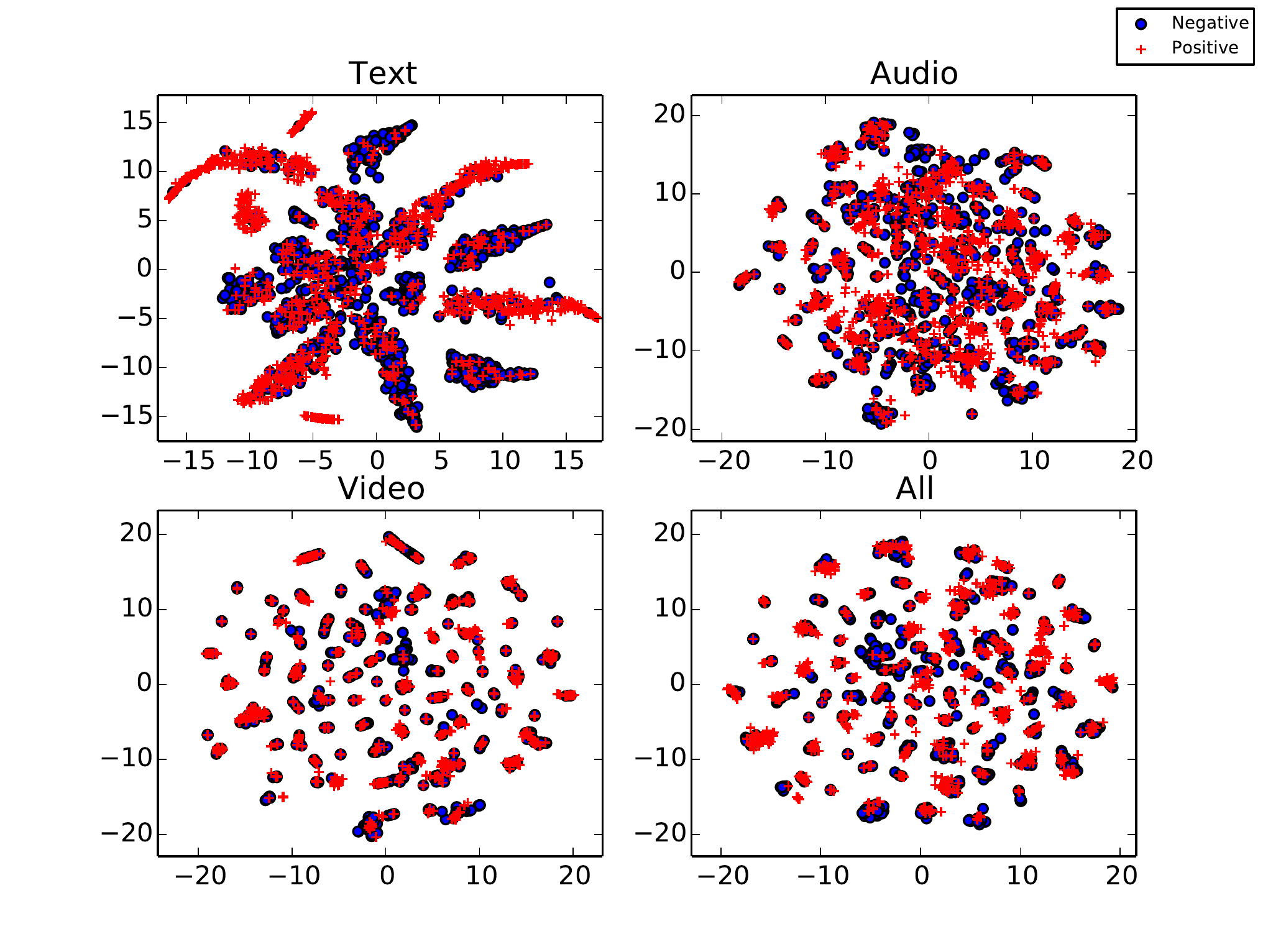}%
	\includegraphics[width=0.5\linewidth]{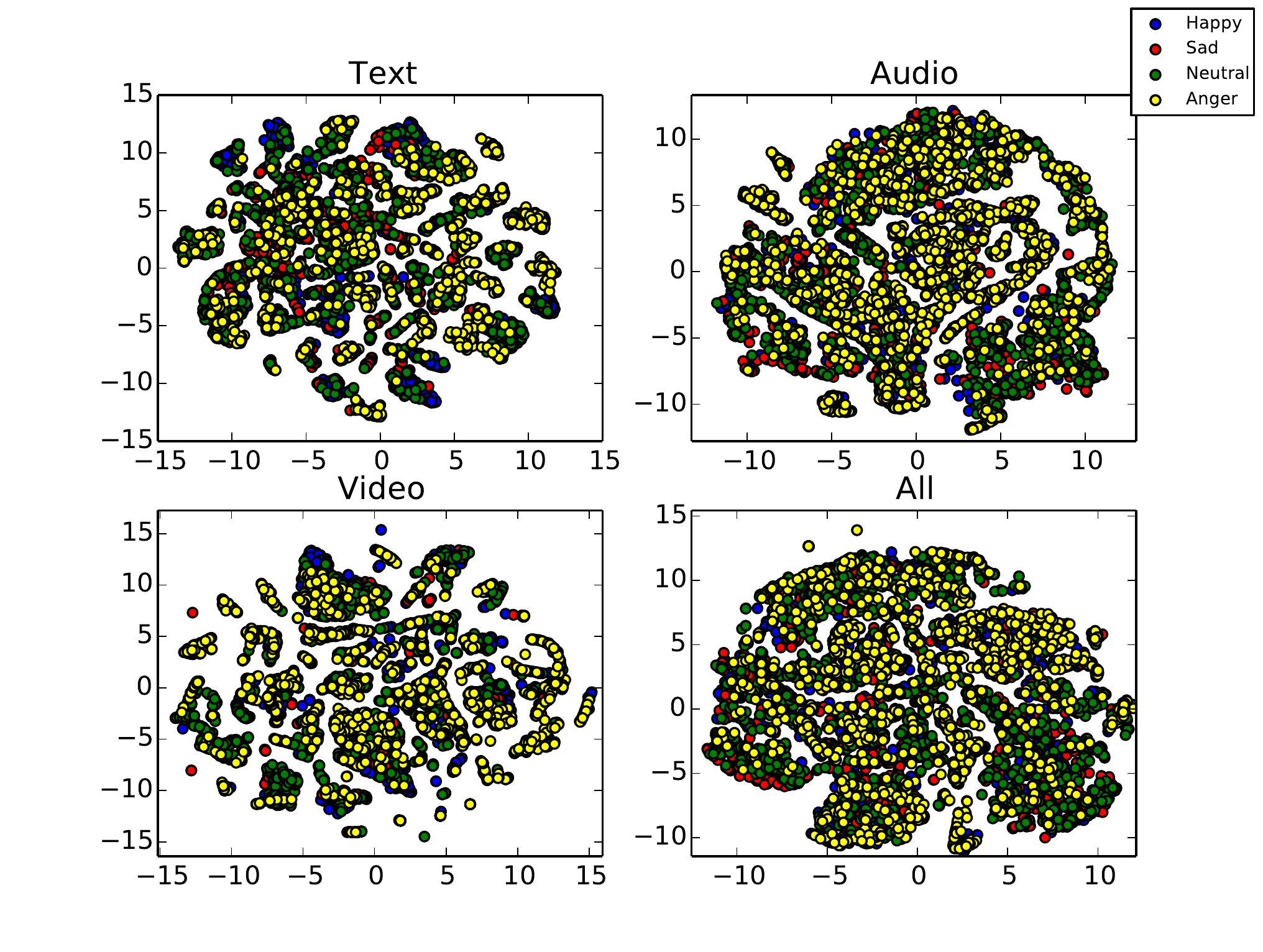}%
	\caption{T-SNE 2D visualization of MOSI and IEMOCAP datasets when unimodal features and multimodal features are used.}
	\label{fig:fig3}
\end{figure}

The MOSI visuals (see Figure \ref{fig:fig3}) present information regarding dataset distribution within single and multiple modalities. For the textual and audio modalities, comprehensive clustering can be seen with substantial overlap. However, this problem is reduced in the video and all modalities with structured de\-clustering but overlap is reduced only in multimodal. This forms an intuitive explanation of the improved performance in the multi\-modality. 

The IEMOCAP visualizations (see Figure \ref{fig:fig3}) provide insight for the 4 class distribution for uni and multimodals, where clearly, the multi\-modal distribution has the least overlap (increase in red and blue visuals, apart from the rest) with sparse distribution aiding the classification process.

\section{Qualitative Analysis}
\label{qa}
In order to have a better understanding on roles of modalities for overall classification, we have manually done some qualitative analysis. Here we show the cases where our model successfully comprehends the semantics of the utterances and with aid from the multiple media, correctly classifies the emotion of the same.

While over-viewing the correctly classified utterances in the validation set, we found that text modality often helped classification of utterances where visual and audio cues are flat with less variance. The model, in such situations, gathered information from the language semantics extracted by the text modality. For example, in an utterance from the MOSI dataset - "amazing special effects",  there was no jest of enthusiasm in speaker's voice and face audio-visual classifier, which caused failure to identify the positivity of this utterance by the audio and video unimodal classifiers. On the other textual classifier given the presence of highly polar words, correctly detected the polarity as positive and helped the bi and multimodal classifiers for correct classification. 

Text modality also helped in situations where the face of the reviewer was not prominent. This result is promising since in many reviews, often the video diverges from the face of the reviewer to other images of products or references. 

However, in some utterances, text modality misclassified due to the presence of misleading linguistic cues. But, the overall classification was correct given the indicative hints from the audio and video inputs. For example, textual classifier classified this sentence - "that like to see comic book characters treated responsibly" as positive, possibly because of the presence of positive phrases such as "like to see", "responsibly". However, the high pitch of anger in the person's voice and the frowning face helps identify this to be a negative utterance.

The above examples demonstrates the effectiveness and robustness of our model to capture overall video semantics of the utterances for emotion and sentiment detection. It also shows how bi and multimodal models, given the multiple media input, overcomes the limitations of  unimodal networks. 

We also explored the misclassified validation utterances and found some interesting trends. A video is constituent of a group of utterances which have contextual dependencies among them. Thus, our model failed to classify utterances whose emotional polarity was highly dependent on the context described in earlier or later part of the video. However, such interdependent modeling was out of the scope of this paper and we therefore enlist it as a future work.

\section{Conclusion}
\label{sec:conclusion}
We have presented a framework for multimodal sentiment analysis and multimodal emotion recognition, which outperforms the state of the art in both tasks by a significant margin. Apart from that, we also discuss some major aspects of multimodal sentiment analysis problem such as the performance of speaker-independent models and cross dataset performance of the models.

Our future work will focus on extracting semantics from the visual features, relatedness of the cross-modal features and their fusion. We will also include contextual dependency learning in our model to overcome limitations mentioned in Section~\ref{qa}. Our framework is available as a demo on \url{http://148.204.64.164/.}\footnote{Best viewed in Mozilla Firefox}
\bibliographystyle{splncs}
\bibliography{paper}

\begin{thebibliography}{10}

\bibitem{perez2013utterance}
P{\'e}rez-Rosas, V., Mihalcea, R., Morency, L.P.:
\newblock Utterance-level multimodal sentiment analysis.
\newblock In: ACL (1). (2013)  973--982

\bibitem{wollmer2013youtube}
Wollmer, M., Weninger, F., Knaup, T., Schuller, B., Sun, C., Sagae, K.,
  Morency, L.P.:
\newblock Youtube movie reviews: Sentiment analysis in an audio-visual context.
\newblock IEEE Intelligent Systems \textbf{28} (2013)  46--53

\bibitem{poria2015deep}
Poria, S., Cambria, E., Gelbukh, A.:
\newblock Deep convolutional neural network textual features and multiple
  kernel learning for utterance-level multimodal sentiment analysis.
\newblock In: Proceedings of EMNLP. (2015)  2539--2544

\bibitem{zadeh2015micro}
Zadeh, A.:
\newblock Micro-opinion sentiment intensity analysis and summarization in
  online videos.
\newblock In: Proceedings of the 2015 ACM on International Conference on
  Multimodal Interaction, ACM (2015)  587--591

\bibitem{camacsa}
Cambria, E.:
\newblock Affective computing and sentiment analysis.
\newblock IEEE Intelligent Systems \textbf{31} (2016)  102--107

\bibitem{pang2002thumbs}
Pang, B., Lee, L., Vaithyanathan, S.:
\newblock Thumbs up?: sentiment classification using machine learning
  techniques.
\newblock In: Proceedings of ACL, Association for Computational Linguistics
  (2002)  79--86

\bibitem{socher2013recursive}
Socher, R., Perelygin, A., Wu, J.Y., Chuang, J., Manning, C.D., Ng, A.Y.,
  Potts, C.:
\newblock Recursive deep models for semantic compositionality over a sentiment
  treebank.
\newblock In: Proceedings of EMNL). Volume 1631., Citeseer (2013)  1642

\bibitem{ekman1974universal}
Ekman, P.:
\newblock Universal facial expressions of emotion.
\newblock Culture and Personality: Contemporary Readings/Chicago (1974)

\bibitem{datcu2008semantic}
Datcu, D., Rothkrantz, L.:
\newblock Semantic audio-visual data fusion for automatic emotion recognition.
\newblock Euromedia'2008 (2008)

\bibitem{de1997facial}
De~Silva, L.C., Miyasato, T., Nakatsu, R.:
\newblock Facial emotion recognition using multi-modal information.
\newblock In: Proceedings of ICICS. Volume~1., IEEE (1997)  397--401

\bibitem{chen1998multimodal}
Chen, L.S., Huang, T.S., Miyasato, T., Nakatsu, R.:
\newblock Multimodal human emotion/expression recognition.
\newblock In: Proceedings of the Third IEEE International Conference on
  Automatic Face and Gesture Recognition, IEEE (1998)  366--371

\bibitem{kessous2010multimodal}
Kessous, L., Castellano, G., Caridakis, G.:
\newblock Multimodal emotion recognition in speech-based interaction using
  facial expression, body gesture and acoustic analysis.
\newblock Journal on Multimodal User Interfaces \textbf{3} (2010)  33--48

\bibitem{schuller2011recognizing}
Schuller, B.:
\newblock Recognizing affect from linguistic information in 3d continuous
  space.
\newblock IEEE Transactions on Affective Computing \textbf{2} (2011)  192--205

\bibitem{rozgic2012speech}
Rozgic, V., Ananthakrishnan, S., Saleem, S., Kumar, R., Prasad, R.:
\newblock Ensemble of {SVM} trees for multimodal emotion recognition.
\newblock In: Proceedings of APSIPA ASC, IEEE (2012)  1--4

\bibitem{metallinou2008audio}
Metallinou, A., Lee, S., Narayanan, S.:
\newblock Audio-visual emotion recognition using gaussian mixture models for
  face and voice.
\newblock In: Tenth IEEE International Symposium on ISM 2008, IEEE (2008)
  250--257

\bibitem{eyben2010line}
Eyben, F., W{\"o}llmer, M., Graves, A., Schuller, B., Douglas-Cowie, E., Cowie,
  R.:
\newblock On-line emotion recognition in a 3-d activation-valence-time
  continuum using acoustic and linguistic cues.
\newblock Journal on Multimodal User Interfaces \textbf{3} (2010)  7--19

\bibitem{wu2011emotion}
Wu, C.H., Liang, W.B.:
\newblock Emotion recognition of affective speech based on multiple classifiers
  using acoustic-prosodic information and semantic labels.
\newblock IEEE Transactions on Affective Computing \textbf{2} (2011)  10--21

\bibitem{mikolov2013efficient}
Mikolov, T., Chen, K., Corrado, G., Dean, J.:
\newblock Efficient estimation of word representations in vector space.
\newblock arXiv preprint arXiv:1301.3781 (2013)

\bibitem{eyben2010opensmile}
Eyben, F., W{\"o}llmer, M., Schuller, B.:
\newblock Opensmile: the munich versatile and fast open-source audio feature
  extractor.
\newblock In: Proceedings of the 18th ACM international conference on
  Multimedia, ACM (2010)  1459--1462

\bibitem{baltruvsaitis20123d}
Baltru{\v{s}}aitis, T., Robinson, P., Morency, L.P.:
\newblock 3d constrained local model for rigid and non-rigid facial tracking.
\newblock In: Computer Vision and Pattern Recognition (CVPR), IEEE (2012)
  2610--2617

\bibitem{zadeh2016multimodal}
Zadeh, A., Zellers, R., Pincus, E., Morency, L.P.:
\newblock Multimodal sentiment intensity analysis in videos: Facial gestures
  and verbal messages.
\newblock IEEE Intelligent Systems \textbf{31} (2016)  82--88

\bibitem{busso2008iemocap}
Busso, C., Bulut, M., Lee, C.C., Kazemzadeh, A., Mower, E., Kim, S., Chang,
  J.N., Lee, S., Narayanan, S.S.:
\newblock Iemocap: Interactive emotional dyadic motion capture database.
\newblock Language resources and evaluation \textbf{42} (2008)  335--359

\bibitem{rozgic2012ensemble}
Rozgi{\'c}, V., Ananthakrishnan, S., Saleem, S., Kumar, R., Prasad, R.:
\newblock Ensemble of svm trees for multimodal emotion recognition.
\newblock In: Signal \& Information Processing Association Annual Summit and
  Conference (APSIPA ASC), IEEE (2012)  1--4

\end{thebibliography}
\end{document}